# Empirical Evaluation of Data Poisoning Attacks in Supervised Learning

Toshif Khan

Department of Math, Data & Technology

Minot State University

Minot, ND, 58707

toshif.khan@minotstateu.edu

Muhammad Abusaqer

Department of Math, Data & Technology

Minot State University

Minot, ND, 58707

muhammad.abusaqer@minotstateu.edu

Authors' version. Presented at the 58th Midwest Instruction and Computing Symposium (MICS 2026), Eau Claire, WI, March 27 to 28, 2026. Trustworthy Language Intelligence Lab, Minot State University.

## Abstract

Data poisoning corrupts training data to degrade a model or to plant attacker-controlled behavior. This study evaluates two representative training-time attacks, label flipping and backdoor poisoning, on MNIST and Fashion-MNIST with three baseline classifiers: Logistic Regression, Linear SVM, and Random Forest. Clean training is compared with poisoning rates of 5%, 10%, and 20% using clean-test accuracy, macro-precision, macro-recall, macro-F1, and, for backdoors, attack success rate. Label flipping caused clear degradation, largest for Logistic Regression and Linear SVM, while Random Forest stayed comparatively stable. Backdoor poisoning reached attack success rates from 0.9667 to 1.0000 on both datasets and all three models while often keeping clean-test performance near baseline. The results separate indiscriminate poisoning, which shows up in standard metrics, from targeted backdoor poisoning, which stays comparatively stealthy while embedding highly effective malicious behavior, and they support security-oriented evaluation beyond conventional clean-test metrics.

# 1. Introduction

Machine learning systems increasingly support decisions in security-sensitive and safety-relevant settings. Their effectiveness, however, depends heavily on the integrity of the data used during training. When an adversary manipulates the training set, the resulting model may learn corrupted decision boundaries, suffer degraded predictive performance, or exhibit attacker-directed behavior during inference [1], [2]. This threat is commonly studied under the broader area of adversarial machine learning, where poisoning attacks target the training phase rather than the inference phase [1].

Among training-time attacks, two poisoning families are especially important. The first is label flipping, in which selected training instances are assigned incorrect labels to reduce overall model quality [2], [5]. The second is backdoor poisoning, in which training samples are modified with a trigger pattern and relabeled so that the trained model later responds incorrectly whenever the same trigger appears [3], [4]. Label flipping is generally used to study indiscriminate degradation of performance, whereas backdoor poisoning represents a targeted threat in which the model may preserve strong clean accuracy while remaining compromised on attacker-chosen inputs [1], [3].

Data poisoning matters because it undermines one of the most basic assumptions of supervised learning: that the training data is trustworthy and representative of the task distribution. As poisoning research has expanded from classical machine-learning settings to deep learning, federated learning, and large language models, the need for reproducible empirical baselines has become more important [6]–[9]. In particular, there is value in compact benchmark studies that compare representative attack types under a single controlled protocol, especially when the goal is to clarify attack behavior rather than to propose a new algorithm.

This paper presents an empirical study of data poisoning using two benchmark image-classification datasets, MNIST and Fashion-MNIST, and three widely used baseline models: Logistic Regression, Linear Support Vector Machine, and Random Forest. The study evaluates the effect of label flipping and backdoor-poisoning attacks at multiple poisoning rates and measures their impact using clean-test classification metrics and, for backdoor attacks, attack success rate. This benchmark-oriented design is consistent with the second author's recent comparative machine-learning studies, which emphasize systematic evaluation across models and metrics rather than claims of methodological novelty [13], [14].

The main contributions of this paper are as follows. First, it provides a unified empirical comparison of indiscriminate and targeted poisoning through label flipping and backdoor triggers. Second, it evaluates how poisoning effects vary across two standard benchmark datasets and three classical baseline classifiers. Third, it offers a concise and reproducible baseline that can help students and practitioners understand poisoning risk, interpret trade-offs among models, and identify the distinct behavior of attacks that degrade global accuracy versus attacks that preserve clean performance while embedding malicious behavior.

## 2. Related Work

Training-data poisoning has been studied for more than a decade and now spans classical machine learning, deep learning, online learning, and large language models. NIST's recent adversarial-machine-learning taxonomy identifies poisoning as a training-time attack class whose goal is to corrupt the learned model through manipulated training data [1]. Recent surveys further show that poisoning research has expanded from early degradation attacks to a broader landscape that includes targeted attacks, backdoors, optimization-based poisoning, federated settings, and LLM-focused contamination studies [6], [7].

Early work on poisoning emphasized the vulnerability of traditional classifiers to corrupted training samples. Biggio et al. showed that support vector machines can be poisoned through carefully crafted training points that increase test error, establishing one of the foundational formulations of poisoning against supervised classifiers [2]. Yerlikaya and Bahtiyar later examined data poisoning against multiple classical machine-learning algorithms, including SVM, Logistic Regression, and Random Forest, with particular attention to random and distance-based label flipping attacks [5]. Their results are especially relevant to the present study because they demonstrate that even relatively simple poisoning strategies can meaningfully reduce performance in classical learning settings.

A major branch of the literature focuses on backdoor poisoning, where the attacker seeks targeted behavior rather than broad accuracy degradation. BadNets remains a landmark study in this area, showing that models trained on poisoned data can behave normally on clean inputs yet misclassify attacker-triggered inputs with high reliability [3]. Schwarzschild et al. later highlighted the importance of standardized evaluation by introducing a unified benchmark for backdoor and targeted poisoning attacks, demonstrating that attack performance can vary substantially across training protocols, architectures, and dataset settings [4]. These findings motivate the use of explicit backdoor metrics, especially attack success rate, rather than relying solely on clean-test accuracy.

Poisoning research has also broadened beyond static offline settings. Zhang et al. formulated online data poisoning as a sequential decision problem, showing that poisoning can be studied in dynamic learning environments rather than only in fixed-batch scenarios [8]. More recently, Zhao et al. surveyed poisoning in deep learning and emphasized the growing importance of attack categorization, scalable attack construction, and new threat surfaces such as large language models [7]. Bowen et al. then examined scaling trends in LLM poisoning, reporting that larger models can become highly susceptible to harmful behavior from relatively small poisoned fractions under some settings [9]. Although such work expands the field significantly, it also increases the gap between advanced poisoning literature and small, reproducible benchmark studies that remain accessible in educational and applied settings.

Taken together, the literature shows three clear patterns. First, poisoning attacks can be either indiscriminate, as in label flipping attacks that reduce overall accuracy, or targeted, as in backdoor attacks that implant trigger-based behavior [1], [6], [7]. Second, prior work has shown that performance must be evaluated carefully because clean accuracy alone may hide successful compromise [3], [4]. Third, while many recent studies focus on deep architectures, complex optimization, or LLM-scale settings [7]–[9], there remains value in concise experimental studies that compare representative poisoning attacks under a common protocol using standard benchmarks and interpretable baseline models. The present paper addresses this need by comparing label flipping and backdoor poisoning on MNIST and Fashion-MNIST using Logistic Regression, Linear SVM, and Random Forest, thereby offering a streamlined but informative empirical baseline.

# 3. Methodology and Experimental Setup

## 3.1 Research Design

This study adopts a controlled benchmark-based design to evaluate how training-time data poisoning affects supervised image classification. The scope is intentionally empirical. The paper does not propose a new poisoning algorithm or a new defense; rather, it examines two representative poisoning types under a unified protocol: label flipping as an indiscriminate poisoning attack and backdoor poisoning as a targeted poisoning attack [1]–[7]. The objective is to measure how these attacks influence model behavior across different datasets, poisoning rates, and baseline classifiers.

## 3.2 Datasets

Two benchmark image-classification datasets were used: MNIST and Fashion-MNIST. MNIST is a standard handwritten-digit benchmark introduced by LeCun et al. [10]. Fashion-MNIST was proposed as a more challenging drop-in replacement for MNIST while preserving the same image size, train-test structure, and number of classes [11]. Both datasets contain 28 × 28 grayscale images distributed across 10 classes [10], [11].

For this study, both datasets were accessed through the scikit-learn and OpenML pipeline [12]. To keep the experiments computationally manageable while preserving class balance, a stratified subset was sampled from each dataset using a fixed random seed of 42. For each dataset, 8,000 samples were used for training and 2,000 for testing. Pixel intensities were normalized to the range [0, 1], and each image was flattened into a 784-dimensional feature vector before model training.

## 3.3 Baseline Models

Three baseline classifiers were selected: Logistic Regression, Linear Support Vector Machine, and Random Forest. These models were chosen because they are well established, computationally efficient, and suitable for transparent benchmark comparison. The selection is also consistent with the comparative evaluation style used in the second author’s recent machine-learning studies, where different model families are assessed under a common metric framework [13], [14].

Logistic Regression and Linear SVM were implemented in pipelines that included feature standardization. The Logistic Regression classifier used the saga solver with a maximum of 300 iterations. The Linear SVM used a maximum of 5,000 iterations. The Random Forest classifier was configured with 100 trees. All models were implemented in scikit-learn [12].

## 3.4 Label Flipping Attack

The first attack scenario models label corruption. For each dataset, poisoning rates of 5%, 10%, and 20% were examined. At a given poisoning rate, the corresponding fraction of training samples was selected uniformly at random. The true label of each selected sample was then replaced with a randomly chosen incorrect label, while the image content itself remained unchanged. This attack follows the general formulation of label-based poisoning in which the adversary corrupts supervision to degrade the learned classifier [2], [5]. The purpose of this experiment was to evaluate how quickly clean-test performance deteriorates as the proportion of corrupted labels increases.

## 3.5 Backdoor-Poisoning Attack

The second attack scenario models trigger-based backdoor poisoning. Following the general BadNets framework, a simple and transparent trigger-stamping procedure was used [3]. For each poisoning rate of 5%, 10%, and 20%, a subset of non-target training samples was selected at random. A visible $3 \times 3$ white square was inserted into the bottom-right corner of each selected image, and the label of the modified image was reassigned to a fixed target class. In all experiments, the target class was set to class 0 for consistency. This process creates a poisoned training set in which the trigger becomes spuriously associated with the target label. The purpose of this experiment was to determine whether a model could maintain relatively strong clean accuracy while simultaneously learning a malicious trigger-based mapping [3], [4].

## 3.6 Training Protocol

For every dataset-model combination, a clean baseline was first trained on the unmodified training set. Each model was then retrained separately under every poisoning condition. Poisoning was applied only to the training data. The clean test set remained unmodified for standard classification evaluation. In the backdoor setting, a second test condition was created by stamping the trigger onto non-target test samples in order to measure targeted attack behavior.

All experiments were implemented in Python using NumPy, pandas, matplotlib, and scikit-learn [12]. A fixed random seed was used throughout the pipeline to improve reproducibility. The experimental notebook was designed to save summary tables and plots automatically for subsequent integration into the Results section.

### 3.7 Evaluation Metrics

Model performance on clean test data was measured using accuracy, macro-precision, macro-recall, and macro-F1. These metrics provide a balanced evaluation of multiclass predictive quality and are widely used in comparative classification studies [5], [13], [14]. In addition, the backdoor setting used attack success rate as a targeted-security metric. To compute this value, the backdoor trigger was applied to test samples whose true class was not the target class, and the fraction of those triggered samples classified as the target class was recorded. This metric captures how effectively the backdoor objective was learned, independently of standard clean-test accuracy [3], [4].

Because the final scope of this paper is centered on attack impact rather than defense benchmarking, the core experimental design emphasizes the comparison of attack type, poisoning rate, dataset, and baseline classifier under one consistent methodology.

## 4. Results

This section reports the empirical results for the final scope of the study, which focuses on two poisoning attacks, label flipping and backdoor poisoning, evaluated on MNIST and Fashion-MNIST with Logistic Regression, Linear SVM, and Random Forest.

### 4.1 Clean Baseline Performance

Table 1 summarizes clean baseline performance before any poisoning was applied. Across both datasets, Random Forest produced the strongest baseline results, while Linear SVM produced the weakest. MNIST was consistently easier than Fashion-MNIST for all three models. On MNIST, accuracy and macro-F1 ranged from 0.8400/0.8358 for Linear SVM to 0.9475/0.9470 for Random Forest. On Fashion-MNIST, the corresponding ranges were 0.7665/0.7656 to 0.8455/0.8435. These clean baselines provide the reference point for interpreting the impact of poisoning in the later experiments.

| Dataset | Model | Accuracy | Macro-F1 |
|---|---|---|---|
| MNIST | Logistic Regression | 0.9005 | 0.8985 |
| MNIST | Linear SVM | 0.8400 | 0.8358 |
| MNIST | Random Forest | 0.9475 | 0.9470 |
| Fashion-MNIST | Logistic Regression | 0.8295 | 0.8287 |
| Fashion-MNIST | Linear SVM | 0.7665 | 0.7656 |
| Fashion-MNIST | Random Forest | 0.8455 | 0.8435 |

Table 1: Clean baseline performance on MNIST and Fashion-MNIST.

## 4.2 Label Flipping Results

Table 2 and Figures 1 and 2 summarize the effect of label flipping on clean-test performance. As the poisoning rate increased from 5% to 20%, the degradation in both accuracy and macro-F1 was most visible for Logistic Regression and Linear SVM, whereas Random Forest remained comparatively stable. Because macro-F1 followed the same overall direction as accuracy, the results indicate that label corruption broadly degraded predictive quality rather than affecting only a small subset of classes.

| Dataset | Poison Rate | Model | Accuracy | Acc. Drop | Macro-F1 | F1 Drop |
|---|---|---|---|---|---|---|
| MNIST | 5% | Logistic Regression | 0.8765 | 0.0240 | 0.8741 | 0.0244 |
| MNIST | 5% | Linear SVM | 0.8125 | 0.0275 | 0.8087 | 0.0271 |
| MNIST | 5% | Random Forest | 0.9420 | 0.0055 | 0.9414 | 0.0056 |
| MNIST | 10% | Logistic Regression | 0.8590 | 0.0415 | 0.8566 | 0.0418 |
| MNIST | 10% | Linear SVM | 0.8045 | 0.0355 | 0.8009 | 0.0349 |
| MNIST | 10% | Random Forest | 0.9440 | 0.0035 | 0.9434 | 0.0035 |
| MNIST | 20% | Logistic Regression | 0.8300 | 0.0705 | 0.8269 | 0.0716 |
| MNIST | 20% | Linear SVM | 0.7905 | 0.0495 | 0.7866 | 0.0492 |
| MNIST | 20% | Random Forest | 0.9400 | 0.0075 | 0.9393 | 0.0077 |
| Fashion-MNIST | 5% | Logistic Regression | 0.8105 | 0.0190 | 0.8092 | 0.0195 |
| Fashion-MNIST | 5% | Linear SVM | 0.7150 | 0.0515 | 0.7123 | 0.0533 |
| Fashion-MNIST | 5% | Random Forest | 0.8385 | 0.0070 | 0.8364 | 0.0070 |
| Fashion-MNIST | 10% | Logistic Regression | 0.7975 | 0.0320 | 0.7972 | 0.0315 |
| Fashion-MNIST | 10% | Linear SVM | 0.7005 | 0.0660 | 0.6975 | 0.0680 |

| Dataset | Poison Rate | Model | Accuracy | Acc. Drop | Macro-F1 | F1 Drop |
|---|---|---|---|---|---|---|
| Fashion-MNIST | 10% | Random Forest | 0.8395 | 0.0060 | 0.8379 | 0.0056 |
| Fashion-MNIST | 20% | Logistic Regression | 0.7795 | 0.0500 | 0.7793 | 0.0494 |
| Fashion-MNIST | 20% | Linear SVM | 0.7065 | 0.0600 | 0.7040 | 0.0616 |
| Fashion-MNIST | 20% | Random Forest | 0.8435 | 0.0020 | 0.8418 | 0.0017 |

Table 2: Label flipping results on MNIST and Fashion-MNIST.

On MNIST, the strongest degradation was observed for Logistic Regression, whose accuracy and macro-F1 dropped by 0.0705 and 0.0716, respectively, at 20% poisoning. Linear SVM also declined noticeably, though less sharply, while Random Forest remained close to baseline across all three poisoning levels, with accuracy drops no larger than 0.0075.

On Fashion-MNIST, the same qualitative pattern appeared, but Linear SVM was the most affected model. Its worst condition occurred at 10% poisoning, where accuracy and macro-F1 dropped by 0.0660 and 0.0680, respectively. Logistic Regression also showed clear degradation, while Random Forest again changed only modestly, with accuracy drops remaining at or below 0.0070.

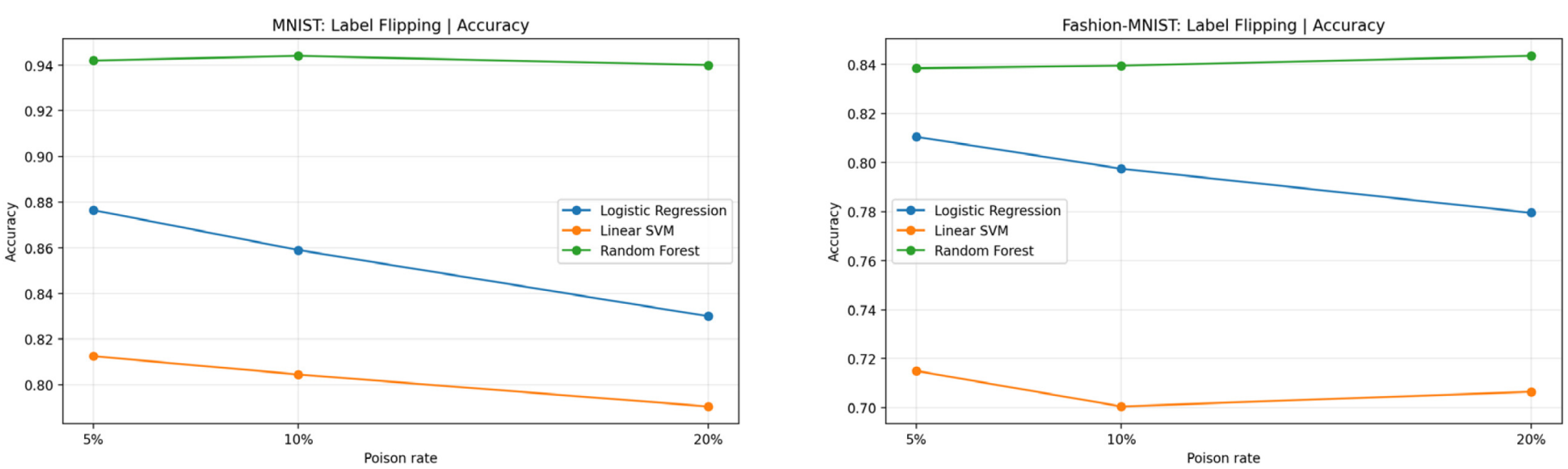

Figure 1: Label flipping accuracy trends on MNIST and Fashion-MNIST.

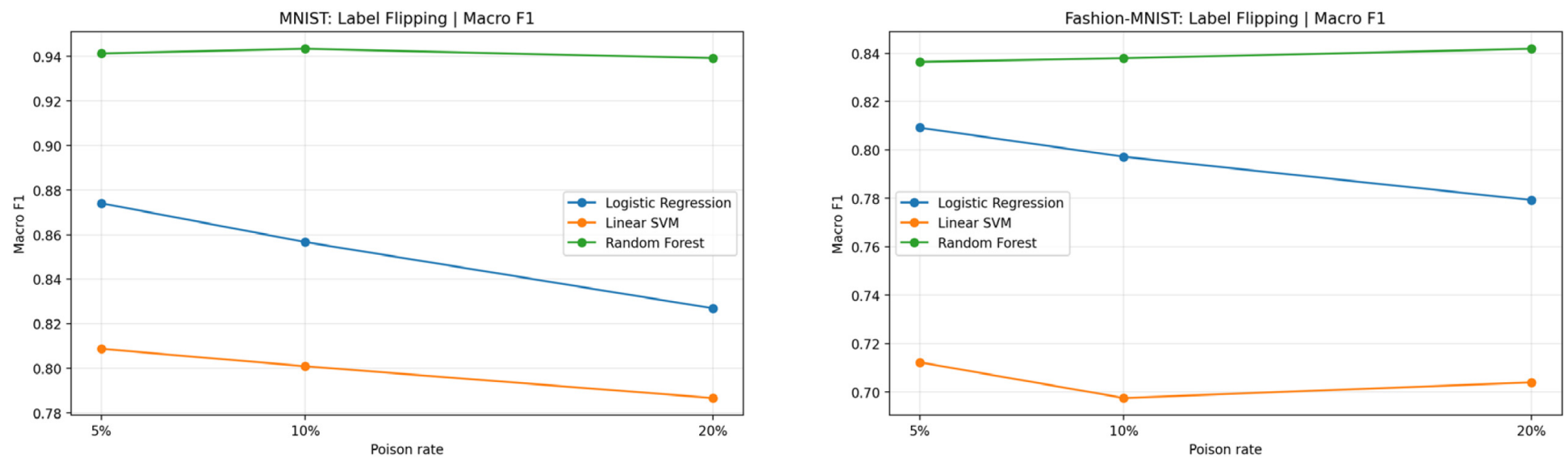

Figure 2: Label flipping macro-F1 trends on MNIST and Fashion-MNIST.

Overall, the label-flipping results show that the two linear models were substantially more sensitive to corrupted supervision than Random Forest in the present experimental setting. In practical terms, label flipping behaved as an indiscriminate attack that produced visible degradation in standard classification metrics.

## 4.3 Backdoor-Poisoning Results

Table 3 and Figures 3 and 4 show the effect of backdoor poisoning. In contrast to label flipping, the backdoor attack produced very high attack success rates while often preserving clean-test accuracy and macro-F1 near baseline levels. This contrast is the most important security result of the study because it shows that a model may appear nearly normal under standard clean-test evaluation while still learning a highly effective trigger-based malicious behavior.

Across all tested backdoor conditions, attack success rate ranged from 0.9667 to 1.0000. On MNIST, changes in clean accuracy remained within 0.0115 across all models and poisoning rates, while attack success rate remained extremely high for every classifier. On Fashion-MNIST, the largest clean-accuracy loss under backdoor poisoning was 0.0330, observed for Linear SVM at 20% poisoning, yet attack success rate still remained above 0.99 for that model.

| Dataset | Poison Rate | Model | Accuracy | Acc. Drop | Macro-F1 | ASR |
|---|---|---|---|---|---|---|
| MNIST | 5% | Logistic Regression | 0.9010 | -0.0005 | 0.8992 | 0.9856 |
| MNIST | 5% | Linear SVM | 0.8330 | 0.0070 | 0.8284 | 0.9961 |
| MNIST | 5% | Random Forest | 0.9480 | -0.0005 | 0.9472 | 1.0000 |
| MNIST | 10% | Logistic Regression | 0.8980 | 0.0025 | 0.8959 | 0.9889 |
| MNIST | 10% | Linear SVM | 0.8415 | -0.0015 | 0.8378 | 0.9956 |

| Dataset | Poison Rate | Model | Accuracy | Acc. Drop | Macro-F1 | ASR |
|---|---|---|---|---|---|---|
| MNIST | 10% | Random Forest | 0.9445 | 0.0030 | 0.9437 | 1.0000 |
| MNIST | 20% | Logistic Regression | 0.8980 | 0.0025 | 0.8960 | 0.9884 |
| MNIST | 20% | Linear SVM | 0.8285 | 0.0115 | 0.8244 | 0.9956 |
| MNIST | 20% | Random Forest | 0.9425 | 0.0050 | 0.9419 | 1.0000 |
| Fashion-MNIST | 5% | Logistic Regression | 0.8230 | 0.0065 | 0.8223 | 0.9667 |
| Fashion-MNIST | 5% | Linear SVM | 0.7615 | 0.0050 | 0.7612 | 0.9928 |
| Fashion-MNIST | 5% | Random Forest | 0.8455 | 0.0000 | 0.8431 | 0.9922 |
| Fashion-MNIST | 10% | Logistic Regression | 0.8250 | 0.0045 | 0.8240 | 0.9883 |
| Fashion-MNIST | 10% | Linear SVM | 0.7640 | 0.0025 | 0.7631 | 0.9950 |
| Fashion-MNIST | 10% | Random Forest | 0.8385 | 0.0070 | 0.8365 | 1.0000 |
| Fashion-MNIST | 20% | Logistic Regression | 0.8200 | 0.0095 | 0.8194 | 0.9867 |
| Fashion-MNIST | 20% | Linear SVM | 0.7335 | 0.0330 | 0.7336 | 0.9939 |
| Fashion-MNIST | 20% | Random Forest | 0.8445 | 0.0010 | 0.8413 | 1.0000 |

Table 3: Backdoor-poisoning results on MNIST and Fashion-MNIST.

Random Forest was the most striking case. It achieved the strongest clean baseline performance on both datasets and also showed perfect or near-perfect backdoor success. On MNIST, Random Forest reached an attack success rate of 1.0000 at all three poisoning rates while maintaining clean accuracy close to baseline. On Fashion-MNIST, it reached 1.0000 at 10% and 20% poisoning while again preserving strong clean performance.

These results indicate that standard clean-test evaluation alone is insufficient for detecting this type of compromise. Unlike label flipping, which visibly degrades ordinary performance, backdoor poisoning can remain comparatively stealthy while embedding highly effective attacker-controlled behavior.

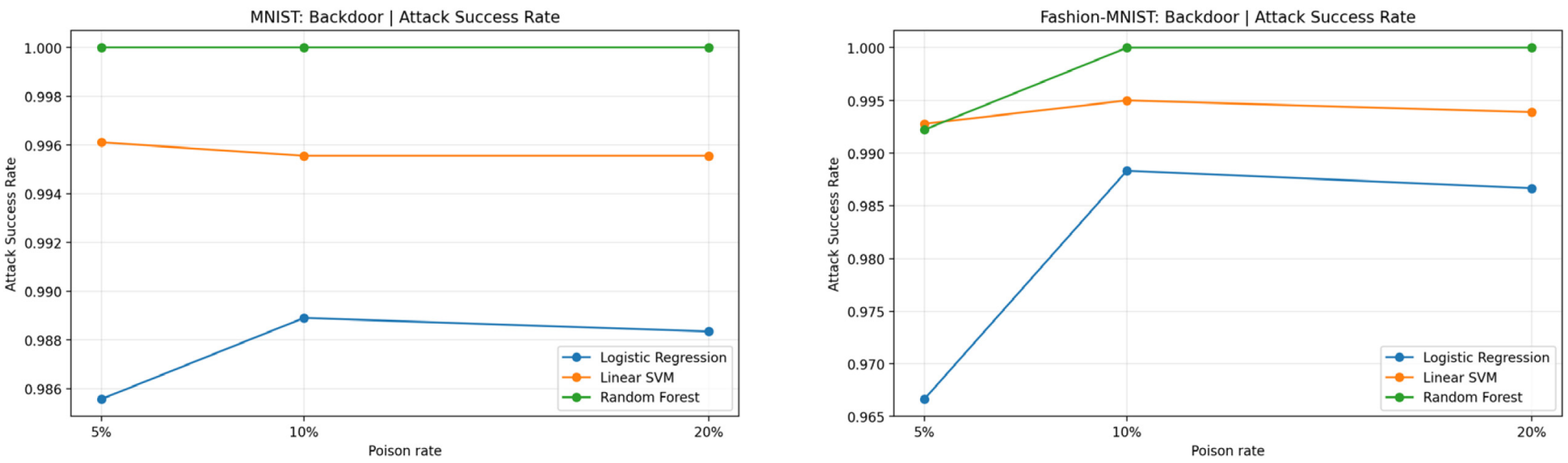

Figure 3: Backdoor attack success rate trends on MNIST and Fashion-MNIST.

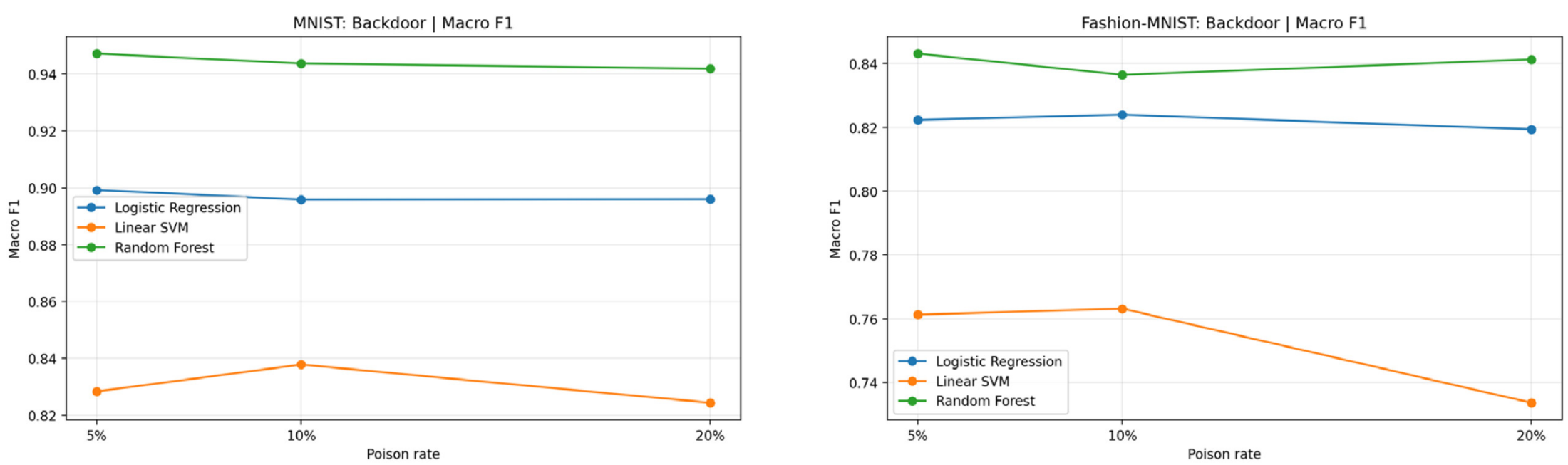

Figure 4: Backdoor macro-F1 trends on MNIST and Fashion-MNIST.

### 4.4 Summary of Findings

Three findings emerge from the experiments. First, MNIST was consistently easier than Fashion-MNIST for all three models, both before and after poisoning. Second, label flipping caused the clearest degradation in clean-test accuracy and macro-F1, with the largest effects observed for Logistic Regression and Linear SVM. Third, backdoor poisoning was the more dangerous attack from a security perspective because it produced near-perfect attack success rates while often preserving clean performance. In other words, label flipping was more visible in standard evaluation metrics, whereas backdoor poisoning was more covert and more effective at implanting attacker-controlled behavior.

## 5. Discussion

The experimental results show a clear distinction between the two poisoning families considered in this paper. Label flipping functioned primarily as an indiscriminate attack that degraded overall predictive quality, whereas backdoor poisoning behaved as a targeted attack that preserved much of the model's ordinary test performance while implanting highly reliable trigger-based misclassification behavior. This contrast is important because it highlights two different failure modes in supervised learning. In the first, the attack is visible through declining standard classification metrics. In the second, the model may appear to perform normally on clean data while remaining substantially compromised from a security perspective [1], [3], [4].

Across both MNIST and Fashion-MNIST, label flipping reduced accuracy and macro-F1 most clearly for Logistic Regression and Linear SVM. These results suggest that the two linear models were more sensitive to corrupted supervision under the present experimental settings. By contrast, Random Forest showed much smaller declines under label flipping on both benchmarks. This does not mean that Random Forest is generally robust to poisoning in all settings; rather, in this study, the ensemble-based model retained stronger clean-test performance under simple random label corruption than the two linear baselines. The result is still useful because it demonstrates that attack impact depends not only on the poisoning rate but also on model family and dataset characteristics.

The backdoor results were more striking. Attack success rate remained extremely high across all three classifiers and both datasets, often approaching or reaching 1.0000, while clean accuracy and macro-F1 remained close to baseline. This pattern is consistent with the broader backdoor literature, which has repeatedly shown that a poisoned model can retain acceptable clean-task behavior while learning a hidden malicious mapping [3], [4]. In practical terms, this means that evaluation based solely on conventional clean-test metrics may fail to reveal a serious compromise. The results therefore reinforce the importance of using targeted security-oriented evaluation measures, such as attack success rate, when studying poisoning threats.

Another meaningful finding is that stronger clean performance did not imply stronger security. Random Forest achieved the best clean baseline results on both datasets, yet it also exhibited perfect or near-perfect backdoor attack success under several settings. This outcome is valuable because it discourages the common but misleading assumption that a high-performing classifier is necessarily more secure. In the present study, the model that performed best under normal conditions also proved highly capable of memorizing the attacker’s trigger. As a result, model quality and model trustworthiness should be treated as related but distinct concerns.

The benchmark choice also influenced the observed outcomes. MNIST was consistently easier than Fashion-MNIST for all three models, both in the clean baseline and under poisoning. This was expected, since Fashion-MNIST is generally regarded as the more challenging classification benchmark [11]. Even so, the same qualitative attack trends appeared on both datasets: label flipping produced visible degradation in standard metrics, while backdoor poisoning achieved high targeted effectiveness with comparatively limited damage to ordinary clean-test performance. This consistency across two tasks strengthens the empirical reliability of the study.

The present findings should also be interpreted with appropriate scope. The experiments were intentionally designed as a compact and reproducible benchmark study rather than a comprehensive adversarial evaluation. The models are classical baselines rather than deep neural architectures, the trigger is simple and visible, and the poisoning rates are moderate. Within those limits, however, the results clearly demonstrate that even straightforward poisoning mechanisms can produce meaningful degradation or highly successful hidden compromise. For a conference paper centered on an accessible empirical baseline, this is an appropriate and informative outcome.

## 6. Future Work

Several directions can extend this study. First, future work can evaluate additional poisoning strategies beyond the two representative attacks used here. Examples include clean-label poisoning, optimization-based poisoning, and more adaptive backdoor triggers. Such extensions would help determine whether the main trends observed in this paper persist under more subtle or more powerful attack formulations [6]–[9].

Second, the experimental scope can be expanded to include more complex model families. The present study focused on Logistic Regression, Linear SVM, and Random Forest because they provide a transparent and computationally manageable baseline. Future research can compare these results against convolutional neural networks, vision transformers, or other deep-learning architectures to examine whether stronger representational capacity changes poisoning susceptibility or attack stealth.

Third, future work should investigate defenses more systematically. Although the broader poisoning literature includes data sanitization, robust training, influence-based filtering, and other mitigation strategies [1], [6], the current paper intentionally prioritized the attack comparison itself. A natural next step is therefore to evaluate stronger defense mechanisms under the same benchmark protocol and to measure the trade-off between mitigation effectiveness and clean-task performance.

Fourth, the study can be extended to more realistic data settings and threat models. Additional datasets, class-imbalance conditions, partial attacker knowledge, and different trigger placements could improve the practical realism of the evaluation. Likewise, future work can examine whether poisoning remains equally effective when the adversary controls only a limited portion of the training pipeline or when the defender uses data validation procedures before training.

Finally, another promising direction is to connect this benchmark-style study to the broader problem of trustworthy artificial intelligence. As poisoning research continues to expand into deep-learning systems, federated learning, and large language models [7]–[9], there is value in developing instructional and experimental frameworks that help students and practitioners understand how security weaknesses can persist even when ordinary performance metrics appear strong. This paper provides one such baseline, and future work can build on it by broadening the task scope, attack realism, and defense depth.

## 7. Conclusion

This paper presented a benchmark-based empirical study of training-time data poisoning using two representative attack types: label flipping and backdoor poisoning. Experiments were conducted on MNIST and Fashion-MNIST using Logistic Regression, Linear SVM, and Random Forest. The results showed that label flipping reduced clean-test performance, especially for the two linear models, while backdoor poisoning achieved very high attack success rates across both datasets and all three classifiers.

The study's most important finding is that poisoning impact depends strongly on the attack objective. Label flipping was more visible through declines in ordinary accuracy and macro-F1, whereas backdoor poisoning was more covert and more dangerous because it preserved much of the clean-task behavior while embedding attacker-controlled misclassification. These results reinforce the idea that conventional evaluation metrics alone are not sufficient for assessing model security in adversarial settings.

Overall, the findings confirm that even simple poisoning strategies can meaningfully affect model trustworthiness under a controlled and reproducible benchmark setting. Although the paper does not propose a new attack or defense, it offers a clear comparative baseline that can support future classroom use, further experimentation, and subsequent studies on poisoning resilience in machine learning systems.